\documentclass[conference]{IEEEtran}
\IEEEoverridecommandlockouts

\usepackage{cite}
\usepackage{amsmath,amssymb,amsfonts}
\usepackage{algorithmic}
\usepackage{graphicx}
\usepackage{textcomp}
\usepackage{xcolor}
\def\BibTeX{{\rm B\kern-.05em{\sc i\kern-.025em b}\kern-.08em
    T\kern-.1667em\lower.7ex\hbox{E}\kern-.125emX}}

\usepackage{multirow}
\usepackage[ruled,vlined]{algorithm2e}
\usepackage{threeparttable}
\usepackage{mathtools}
\usepackage{amsthm}

\DeclarePairedDelimiter{\floor}{\lfloor}{\rfloor}
\theoremstyle{definition}

\newtheorem{innercustomthm}{Theorem}
\newenvironment{customthm}[1]
  {\renewcommand\theinnercustomthm{#1}\innercustomthm}
  {\endinnercustomthm}

\begin{document}

\title{Computing-In-Memory Dataflow for Minimal Buffer Traffic}

\author{Choongseok Song$^{\dag}$ and Doo Seok Jeong$^{\dag\ddag*}$\\
$^{\dag}$Division of Materials Science and Engineering, Hanyang University, Seoul, Republic of Korea\\
$^\ddag$Department of Semiconductor Engineering, Hanyang University, Seoul, Republic of Korea\\
\IEEEauthorblockA{*Corresponding author: $\text{dooseokj}@\text{hanyang.ac.kr}$
}}


\maketitle

\begin{abstract}
Computing-In-Memory (CIM) offers a potential solution to the memory wall issue and can achieve high energy efficiency by minimizing data movement, making it a promising architecture for edge AI devices. Lightweight models like MobileNet and EfficientNet, which utilize depthwise convolution for feature extraction, have been developed for these devices. However, CIM macros often face challenges in accelerating depthwise convolution, including underutilization of CIM memory and heavy buffer traffic. The latter, in particular, has been overlooked despite its significant impact on latency and energy consumption.
To address this, we introduce a novel CIM dataflow that significantly reduces buffer traffic by maximizing data reuse and improving memory utilization during depthwise convolution. The proposed dataflow is grounded in solid theoretical principles, fully demonstrated in this paper. When applied to MobileNet and EfficientNet models, our dataflow reduces buffer traffic by 77.4–87.0\%, leading to a total reduction in data traffic energy and latency by 10.1–17.9\% and 15.6–27.8\%, respectively, compared to the baseline (conventional weight-stationary dataflow).
\end{abstract}


\section{Introduction}
Convolutional neural networks (CNNs) have achieved remarkable success in computer vision, excelling in spatial feature extraction~\cite{he2016resnet}. As CNNs evolve, they tend to grow deeper and larger, requiring more parameters and computational power. However, this growth presents challenges for implementation in edge artificial intelligence (AI) devices~\cite{bianco2018cost}. To address these challenges, lightweight neural network models and low-power hardware platforms, as alternatives to traditional graphics processing units (GPUs), have been developed.

Various strategies to achieve lightweight network models have emerged. They include replacements of conventional convolution operations by lightweight depthwise separable convolution operations as for MobileNets~\cite{howard2017mobilenetv1, sandler2018mobilenetv2, howard2019mobilenetv3}, EfficientNets~\cite{tan2019efficientnet}, and ShuffleNets~\cite{zhang2018shufflenet}. MobileNetV1 achieves high classification accuracy while largely reducing its parameter number (and thus the operation number) by using depthwise separable convolution that combines depthwise convolutions (DWConvs) and pointwise convolutions (PWConvs)~\cite{howard2017mobilenetv1}. MobileNetV2 further reduces the parameter and operation numbers by introducing a bottleneck which is an inverted residual block including depthwise separable convolution~\cite{sandler2018mobilenetv2}.

GPUs, while powerful, suffer from low power-efficiency due to heavy data movement within the memory hierarchy~\cite{song2024hardwarereview}. This is because a vast amount of operands are moved to arithmetic logic units per instruction. Computing-in-memory (CIM) may offer a low-power solution to convolution acceleration by computing within memory, where weights or input activations (IAs) are pre-stored, i.e., stationary in the memory. Because one type of operand is stationary, CIM likely mitigates the data traffic. Further, CIM harnesses the analog computing based on charge accumulation~\cite{lee2022chargepim} or current summation~\cite{dong2020currentpim} in parallel, which likely boosts operational efficiency.

Albeit efficient, a downside is the inherently limited flexibility of data arrangement in the memory such that a weight (or IA) vector subject to operations should be aligned column-wise to share a bitline(s) in the tile memory (TM), and an IA (or weight) vector should be aligned in the tile register file (TRF) accordingly. These constraints leave several crucial challenges for efficient implementations of lightweight network models in CIM macros. They include (1) limited data reuse in CIM macro, and thus heavy buffer and off-chip DRAM traffic and (2) limited utilization of CIM memory, particularly, for DWConv operations. 

To address these issues, we propose a novel CIM dataflow for lightweight network models using a novel data mapping method and scheduling algorithm. Particularly, the proposed dataflow largely reduces the buffer traffic which has been overlooked despite its significant impact on latency and energy consumption. Our primary contributions include:

\begin{itemize} 
\item We introduce a weight-stationary data mapping method (on solid theoretical grounds) that significantly boosts IA reuse (and thus reducing the buffer traffic) by weight duplication and a few step IA shift. 
\item We introduce the BIG/LITTLE scheduler to maintain high memory utilization in CIM macros for various feature map sizes. 
\item We share the in-depth analysis results of our dataflow, considering the contributions of the buffer traffic and data re-mapping to latency and energy consumption, which have been overlooked in previous performance evaluations. 
\end{itemize}

\section{Preliminaries}
\subsection{CIM architecture and buffer traffic}
\begin{figure}[tb]
  \centering
  \includegraphics[width=\linewidth]{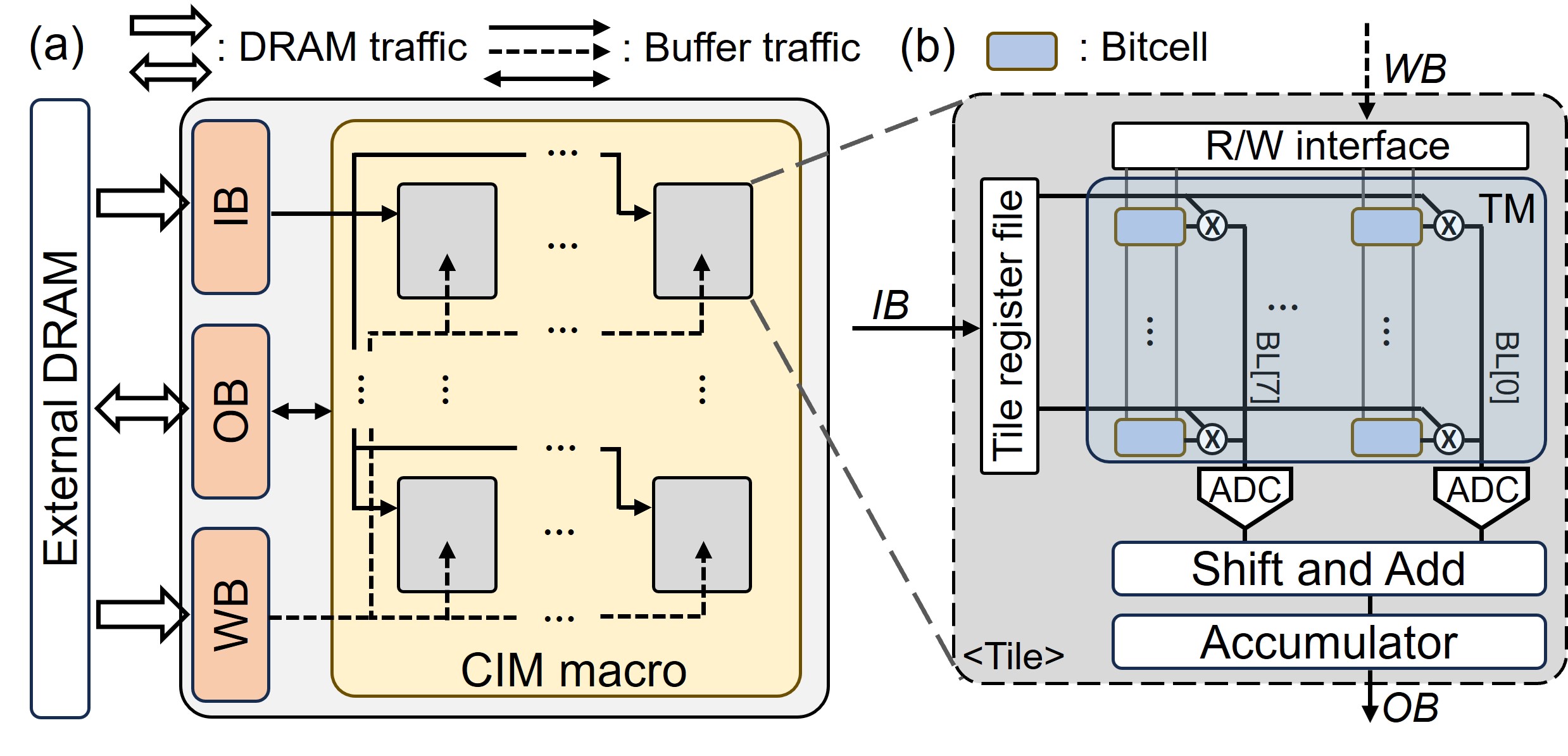}
  \caption{CIM tile architecture and three-level memory hierarchy employed.}
  \label{fig:CIM_system}
\end{figure}

Fig.~\ref{fig:CIM_system} illustrates the CIM tile architecture and memory hierarchy used. The TM ($180\times8$b 8T-SRAM) stores INT8 weights while the TRF stores INT8 IAs, i.e., weight-stationary CIM. Each IA in the TRF is applied to the TM in a bit-serial manner from the LSB to MSB. The inner product result for each bitline is reflected as a voltage drop on the pre-charged bitline, which is then converted to a digital value by an analog-to-digital converter (ADC). These values from eight parallel ADCs are subsequently processed through a bit-shift operation to align with the digits of the IAs and weights and summed in a Shift-and-Add (S\&A) unit. This value is accumulated in an accumulator until the MSB of the IAs is processed.

We consider CIM architecture based on a three-level memory hierarchy: (1) off-chip DRAM, (2) on-chip buffer including input buffer (IB), output buffer (OB), and weight buffer (WB), and (3) TRF and TM in each tile, which directly access IB and WB, respectively. OB buffers partial or completed outputs. While DRAM traffic has been subject to intensive analysis~\cite{lee2024dramtraffic, park2024dramtraffic}, buffer traffic between the buffer and spatially unrolled data across the multiple tiles has been overlooked. For weight-stationary CIM, buffer traffic mainly arises from loading the TRFs with partial input feature maps (ifmaps) from IB. This can be alleviated by reusing the partial ifmaps in the TRFs.

\subsection{Weight- and input-stationary dataflows}\label{sec:dataflow}
In CIM, weight-stationary (WS) dataflows are frequently adopted for 2D convolution (Conv2D) and PWConv such that the weights along the input channel direction are mapped onto each column in the TM to share a bitline. This allows parallel multiply-accumulate (MAC) operations along the input channel direction. Because the ifmaps for Conv2D and PWConv frequently have long input channels, this WS dataflow enables a high utilization rate of the TM. 
However, DWConv needs MAC operations over each kernel height-width plane (typically $3\times 3$ and $5\times 5$) only, so that WS dataflow with mapping such a vectorized kernel height-width plane onto a column in the TM suffers from TM under-utilization. Input-stationary (IS) dataflows can enhance the TM utilization, however, they should frequently re-write the TM (details in Sec.~\ref{sec:access}). Alternatively, there exists a hybrid WS-IS dataflow to address the under-utilization issue for WS dataflows~\cite{lo2023morphablecim}.

\subsection{Convolution with duplicated kernels (ConvDK)}\label{sec:dupkernel_theory}
We introduce a convolution with duplicated kernels (ConvDK) method, which significantly increases an IA reuse rate in CIM tiles with a high TM utilization rate, and thus largely reducing buffer traffic.
\begin{customthm}{1}\label{theo:theorem1}
Let $k$ and $s$ be a positive odd number and positive integer satisfying $s<k$, respectively. Let $m$ and $n$ be non-negative integers that satisfy 
\begin{equation}\label{equ:govern}
ms = nk + a\text{,}
\end{equation}
for a non-negative integer $a<\text{lcm}\left(k,s\right)/s$, where $\text{lcm}\left(k,s\right)$ denotes the least common multiple of $k$ and $s$. We define the least $m$ and $n$ (satisfying Eq.~\eqref{equ:govern} for $a=1$) as $m_1$ and $n_1$, respectively. If $m_1$ and $n_1$ exist, we have
\begin{equation*}
\begin{array}{ll}
&m = i\text{lcm}\left(k,s\right)/s+am_{1} \left(\text{mod }\text{lcm}\left(k,s\right)/s\right)\text{,}\\
&n = j\text{lcm}\left(k,s\right)/k+an_{1} \left(\text{mod }\text{lcm}\left(k,s\right)/k\right)\text{,}
\end{array}
\end{equation*}
where $i$ and $j$ are non-negative integers, and $\text{mod}$ denotes a modulo operation, respectively.
\end{customthm}

\begin{proof}
We define the least $m$ and $n$ satisfying Eq.~\eqref{equ:govern} as $m_{a}$ and $n_{a}$, respectively. Then, Eq.~\eqref{equ:govern} is expressed as
\begin{equation*}
ms = \left(n-n_{a}\right)k + n_{a}k + a = \left(n-n_{a}\right)k + m_{a}s\text{.} 
\end{equation*}
Thus, we have $\left(m-m_{a}\right)s = \left(n-n_{a}\right)k$.
We have the following general solutions for non-negative integers $i$ and $j$.
\begin{equation}\label{equ:general1}
m = i\text{lcm}\left(k,s\right)/s+m_{a}\text{, and } n = j\text{lcm}\left(k,s\right)/k + n_{a}\text{,}
\end{equation}
representing periodicities of $\text{lcm}\left(k,s\right)/s$ and $\text{lcm}\left(k,s\right)/k$, respectively. Since $s<k$, the period of $m$ is longer than $n$. 

Pairs of $m_1$ and $n_1$, and $m_2$ and $n_2$ satisfy 
\begin{equation*}\label{equ:sh12}
m_1s = n_1k+1\text{, and }m_2s = n_2k+2\text{,}
\end{equation*}
respectively. Thus, the following equation holds: 
\begin{equation}\label{equ:diff}
(m_2-m_1)s = (n_2-n_1)k+1\text{.}
\end{equation}
The least $(m_2-m_1)$ and $(n_2-n_1)$ that satisfy Eq.~\eqref{equ:diff} are $m_1$ and $n_1$, and thus we have $m_2 = 2m_1\quad\text{and}\quad n_2 =2n_1$.
A similar analysis applies to general terms $m_{a}$ and $n_{a}$, yielding 
\begin{equation}
\begin{aligned}\label{equ:mn_general}
m_{a} = am_{1}\left(\text{mod }\text{lcm}\left(k,s\right)/s\right)\text{,}\\
n_{a} =an_{1}\left(\text{mod }\text{lcm}\left(k,s\right)/k\right)\text{.}
\end{aligned}
\end{equation}
Note that the modulo operations apply to $am_1$ and $an_1$ because of the periodicity of $m$ and $n$ as shown in Eq.~\eqref{equ:general1}.
By plugging Eq.~\eqref{equ:mn_general} into Eq.~\eqref{equ:general1}, we can calculate $m$ and $n$ for a given $a$. 
\begin{equation*}
\begin{array}{ll}
&m = i\text{lcm}\left(k,s\right)/s + am_{1} \left(\text{mod }\text{lcm}\left(k,s\right)/s\right)\text{,}\\
&n = j\text{lcm}\left(k,s\right)/k + an_{1} \left(\text{mod }\text{lcm}\left(k,s\right)/k\right)\text{.}
\end{array}
\end{equation*}
\end{proof}

\begin{customthm}{2}\label{theo:theorem2}
Let $\boldsymbol{M}_a$ be a set of $m$'s satisfying Eq.~\eqref{equ:govern} for a given $a\in\boldsymbol{A}\left(=\left\{0,1,\cdots,l-1\right\}\right)$, where $l=\text{lcm}\left(k,s\right)/s$. As such, $m_1$ and $n_1$ satisfy $m_1s=n_1k+1$. If $\text{gcd}\left(m_1,l\right)=1$ (gcd denotes the greatest common divisor), the following properties hold.
\begin{equation*}\label{equ:theorem2}
\begin{array}{ll}
&\boldsymbol{M}_a\cap \boldsymbol{M}_{a'}= \emptyset\text{ for $a\neq a'$ and $a$, $a'$}\in\boldsymbol{A}\text{,}\\
&\bigcup_{a\in\boldsymbol{A}}\boldsymbol{M}_a = \left\{m\in\mathbb{Z}|m\geq 0\right\}\text{.}
\end{array}
\end{equation*}
\end{customthm}
\begin{proof}
We first prove the following proposition:\\ 
\textbf{Proposition }1: If $am_1\left(\text{mod }l\right)=a'm_1\left(\text{mod }l\right)$ and $a, a'\in\boldsymbol{A}$, then $a=a'$.\\
To meet this condition, we have $(a-a')m_1 = (b-b')l$, where $b$ and $b'$ are non-negative integers. Thus, we have $a-a' = i\text{lcm}\left(m_1,l\right)/m_1$, where $i$ is a non-negative integer. If $\text{gcd}\left(m_1,l\right)=1$, then $\text{lcm}\left(m_1,l\right)/m_1=l$. Therefore, we have $a-a' = il$. 
Given that $a, a'\in\boldsymbol{A}\left(=\left\{0,1,\cdots,l-1\right\}\right)$, we have $-l+1\leq a - a'=il\leq l-1$. This inequality holds for $i=0$ only, and thus we have $a=a'$. Given \textbf{Proposition }1 is true, its contraposition is also true:\\
\textbf{Proposition }2: If $a\neq a'$ for $a, a'\in\boldsymbol{A}$, then $am_1\left(\text{mod }l\right)\neq a'm_1\left(\text{mod }l\right)$.\\
According to \textbf{Theorem} 1, $\boldsymbol{M}_a=\left\{il+am_{1}\left(\text{mod }l\right)|i\geq0\right\}$. Given \textbf{Proposition} 2, we have $\boldsymbol{M}_a\cap \boldsymbol{M}_{a'}= \emptyset$ for $a\neq a'$. We define a set $\boldsymbol{M}_i=\left\{il+am_{1}\left(\text{mod }l\right)|a\in\boldsymbol{A}\right\}$. Considering \textbf{Proposition }2, we have $\boldsymbol{M}_i=\left\{il, il+1,\cdots,(i+1)l-1\right\}$. Therefore, $\bigcup_{i=0}^{\infty}\boldsymbol{M}_i=\left\{m\in\mathbb{Z}|m\geq 0\right\}$. Because $\bigcup_{a\in\boldsymbol{A}}\boldsymbol{M}_a = \bigcup_{i=0}^{\infty}\boldsymbol{M}_i$, we eventually have $\bigcup_{a\in\boldsymbol{A}}\boldsymbol{M}_a=\left\{m\in\mathbb{Z}|m\geq 0\right\}$.
\end{proof}

\begin{figure}[tb]
  \centering
  \includegraphics[width=\linewidth]{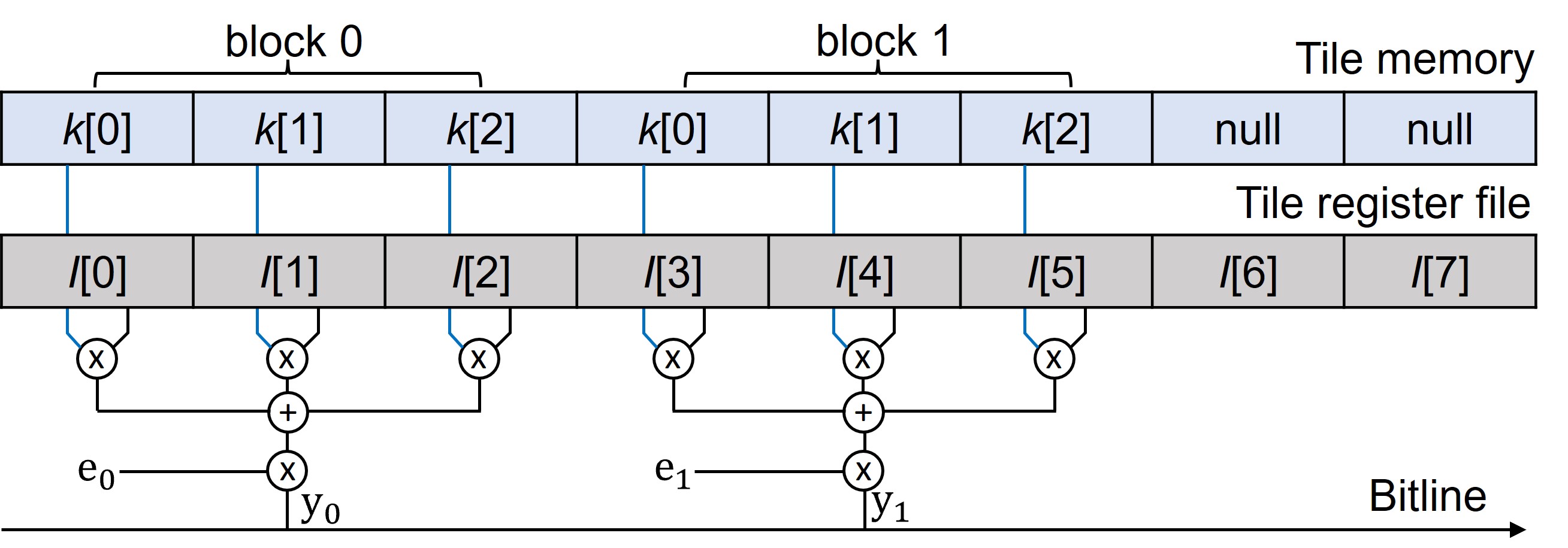}
  \caption{Schematic of 1D ConvDK with a once duplicated 1D kernel ($N=2$, $k_w=3$, $s=1$) for shift-cycle 0. At shift-cycle 1, the tile register file shifts to the left by one.}
  \label{fig:principle}
\end{figure}

For simplicity, let us explain the case of 1D convolution $z=k\ast I$
using a 1D kernel $k$ ($k_w$ in length) with stride $s$. The kernel is $N$-times duplicated, each of which is referred to as a block and indexed by $n$ (Fig.~\ref{fig:principle}). For a better visualization, we represent the TM bitlines with horizontal lines hereafter, so that the TM dimension is expressed as $8\times180$.
The length of the IA vector $I$ is set to $Nk_w+\text{lcm}\left(k_w,s\right)/s-1$. The convolution result from block $n$ is indicated by $y_n$, 
\begin{equation}\label{equ:inner_partial}
  y_n = \sum_{i=0}^{k_w-1}k\left[i\right]I\left[i+nk_w+a\right]\text{,}
\end{equation}
where $a$ denotes a shift parameter for $I$. Eq.~\eqref{equ:inner_partial} corresponds to the element $z\left[m\right]$ that satisfies 
\begin{equation}\label{equ:condition}
  ms = nk_w + a\text{.}
\end{equation}
The maximum of $m$ ($m_\text{max}$) is determined by the kernel duplication number $N$. 
Let us assume that the following conditions on $k_w$ and $s$ for \textbf{Theorems }1 and 2 hold.\\
\textbf{Condition }1: kernel $k_w$ is an odd number, and stride $s$ is smaller than $k_w$.\\
\textbf{Condition }2: There exist non-negative integers $m_1$ and $n_1$ such that $m_1s = n_1k_w + 1$.\\
\textbf{Condition }3: The non-negative integer $m_1$ satisfies $\text{gcd}\left(m_1,l\right)=1$, where $l=\text{lcm}\left(k_w,s\right)/s$.\\
According to \textbf{Theorem} 2, there exists $m(\leq m_\text{max})$ satisfying Eq.~\eqref{equ:condition} for any $a$ in the range $0\leq a\leq \text{lcm}\left(k_w,s\right)/s-1$ if the following equation holds: 
$\text{gcd}\left(m_1,\text{lcm}\left(k_w,s\right)/s\right)=1$. That is, using such a duplicated kernel, the convolution $z=k\ast I$ is completed by shifting vector $I$ by $\text{lcm}\left(k_w,s\right)/s-1$ times. For each $a$, a set of $n$'s such that $e_n=1$ (i.e., multiplication-enable signal is high) and a set of indices $m$'s for the calculated $z[m]$ can be evaluated using \textbf{Theorem} 1. The calculation sequence is elaborated in \textbf{Algorithm}~\ref{algo:conv}. 

\begin{algorithm}[tb]
\DontPrintSemicolon
\SetAlgoLined
\textbf{Input}: IA vector $I\in\mathbb{R}^{Nk_w+\text{lcm}\left(k_w,s\right)/s-1}$\;
\textbf{Output}: $z\left(=k\ast I\right)\in\mathbb{R}^{\floor*{\left((N-1)k_w+\text{lcm}(k_w,s)/s-1\right)/s}+1}$\;
\For{a = 0 \KwTo $\textrm{lcm}\left(k_w,s\right)/s-1$}{
    $n\gets an_1 \left(\text{mod lcm}\left(k_w,s\right)/k_w\right)$\;
    $m\gets am_1 \left(\text{mod lcm}\left(k_w,s\right)/s\right)$\;
    \While{n $<$ N}{
        Calculate $y_n$ using Eq.~\eqref{equ:inner_partial}\;
        $z\left[m\right]\gets y_n$\;
        $n\gets n + \text{lcm}\left(k_w,s\right)/k_w$\;
        $m\gets m + \text{lcm}\left(k_w,s\right)/s$\;}
    }
    \caption{1D ConvDK}
    \label{algo:conv}
\end{algorithm}

\section{Dataflow proposal}
\subsection{Data mapping and parallel operations}\label{sec:datamapping}
\begin{figure}[tb]
  \centering
  \includegraphics[width=\linewidth]{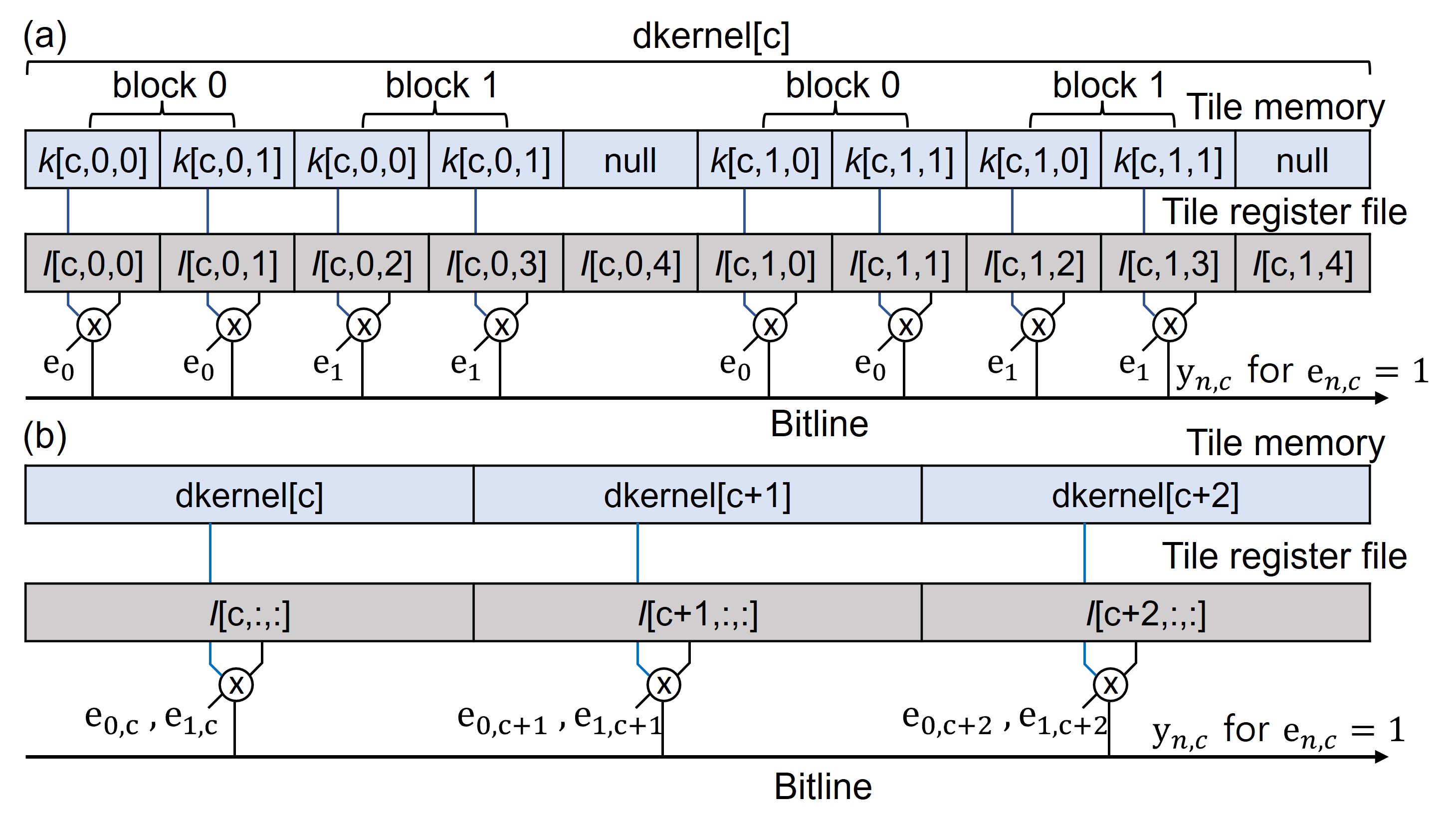}
  \caption{(a) Mapping of IAs to the TRF and weights to TM for Conv2D with a once duplicated 2D kernel ($N=2$, $k_h=k_w=2$, $s=1$) of a given channel $c$ for shift-cycle 0. At shift-cycle 1, the TRF shifts to the left by one. (b) Data mapping for input $I$ and 2D kernels of multiple channels.}
  \label{fig:principle_2d}
\end{figure}

The proposed dataflow utilizes ConvDK that can readily be scaled to higher dimensional kernels and ifmaps. Hereafter we focus on DWConv although the proposed method can easily apply to Conv2D.   
As in Fig.~\ref{fig:principle_2d}(a), the vectorized $k_h\times k_w$ kernel for a specific channel $c$ is duplicated $N$-times and mapped onto the TM. 
Commonly, $k_w$ is an odd number, and stride $s$ is smaller than $k_w$, and thus \textbf{Condition }1 is true. Further, common choices of $k_w$ and $s$ let \textbf{Conditions }2 and 3 be true. Therefore, \textbf{Theorems }1 and 2 are applicable to DWConv. 

A $k_h\times\left[Nk_w+\text{lcm}\left(k_w,s\right)/s-1\right]$ slice of an ifmap serves as an IA vector $I$, which is vectorized and fetched to the $8b\times 180$ TRF (Fig.~\ref{fig:principle_2d}(a)). This sub-map is subject to DWConv across its width.
A set of periodic blocks (indexed by $n$) is endowed with a multiplication-enable signal $e_{n,c}$, where $0\leq n< N$, and $c$ denotes the input(=output) channel index. The result from the tile is given by
\begin{equation}\label{equ:inner_partial2d}
y_{n,a|c,h} = \sum_{i=0}^{k_w-1}\sum_{j=0}^{k_h-1}k\left[c,j,i\right]I\left[c,sh+j,i+nk_w+a\right]\text{.}
\end{equation}
Note that Eq.~\eqref{equ:inner_partial2d} is calculated in parallel in a single tile. DWConv for a channel $c$ is performed in series following \textbf{Algorithm}~\ref{algo:conv} but calculating $y_n$ using Eq.~\eqref{equ:inner_partial2d} instead of Eq.~\eqref{equ:inner_partial}. For instance, for kernel size $k_h=k_w=3$, and stride $s=2$, we have $n_1=1$ and $m_1=2$, for which \textbf{Conditions} 1-3 are true. DWConv thus needs three cycles ($a=0,1,2$) including two IA-shift cycles ($a=1,2$). For duplication number $N=30$, the first cycle ($a=0$, i.e., no IA shift) needs 15 sub-cycles ($n=0, 2, 4, \cdots, 26, 28$) to calculate $z\left[c,h,m\right]$, where $m=0, 3, 6, \cdots, 39, 42$. The second cycle ($a=1$, i.e., IA shift by one) also needs 15 sub-cycles ($n=1, 3, 5, \dots, 27, 29$) for $z\left[c,h,m\right]$, where $m=2, 5, 8, \cdots, 41, 44$. The last cycle ($a=2$, i.e., IA shift by two) also needs 15 sub-cycles ($n=0, 2, 4, \cdots, 26, 28$) for $z\left[c,h,m\right]$, where $m=1, 4, 7, \cdots, 40, 43$, completing DWConv across the width of the sub-map $I$. The kernel is duplicated in the TM, and thus enhancing the TM utilization. Also, the IA loaded in the TRF is reused multiple times, alleviating the buffer traffic.

\begin{algorithm}[tb]
\DontPrintSemicolon
\SetAlgoLined
\For{a = 0 \KwTo $\textrm{lcm}\left(k_w,s\right)/s-1$}{
    Calculate $n$ and $m$ as for \textbf{Algorithm}~\ref{algo:conv}\;
    \While{n $<$ N}{
        \For{$c = 0$ \KwTo $C_0-1$}{
            Calculate $y_{n,a|c,h}$ using Eq.~\eqref{equ:inner_partial2d}\;
            $z\left[c,h,m\right]\gets y_{n,a|c,h}$\;
        }    
        Update $n$ and $m$ as for \textbf{Algorithm}~\ref{algo:conv}\;}
    }
    \caption{DWConv based on ConvDK using multiple 2D kernels of $C_0$ channels across ifmap width.}  
    \label{algo:conv_multichannel}
\end{algorithm}

When the $H\times W$ dimensions of the ifmap are small, multiple channels are loaded into the TRF, enabling the TM to be loaded with 2D ($k_h\times k_w$) kernels from multiple channels. This case is discussed in detail in Sec.~\ref{sec:tiling}. In this case, the TM is organized as in Fig.~\ref{fig:principle_2d}(b). The parallel MAC operations are then iterated over the channel index $c$ for the given values of $a$ and $n$, as described in \textbf{Algorithm}~\ref{algo:conv_multichannel}. 

\subsection{Tiling with BIG/LITTLE scheduler over multiple tiles}\label{sec:tiling}

The proposed dataflow employs a novel BIG/LITTLE scheduler to enhance TM utilization for ConvDK. Depending on the ifmap width, either the BIG or LITTLE schedulers is selected. We define $T_w$ as the largest width of a 2D sub-ifmap (with height $k_h$) that can be fetched in the $8b\times 180$ TRF, where $T_w(=\lfloor180/k_h\rfloor)$. For a $C\times H\times W$ ifmap, the kernel duplication number $N$ within each tile is given by 
\begin{equation}\label{equ:num_dupls}
N=\left(\min\left(W,T_w\right)-\text{lcm}\left(k_w,s\right)/s+1\right)/k_w\text{.}
\end{equation}
\begin{figure}[tb]
  \centering
  \includegraphics[width=\linewidth]{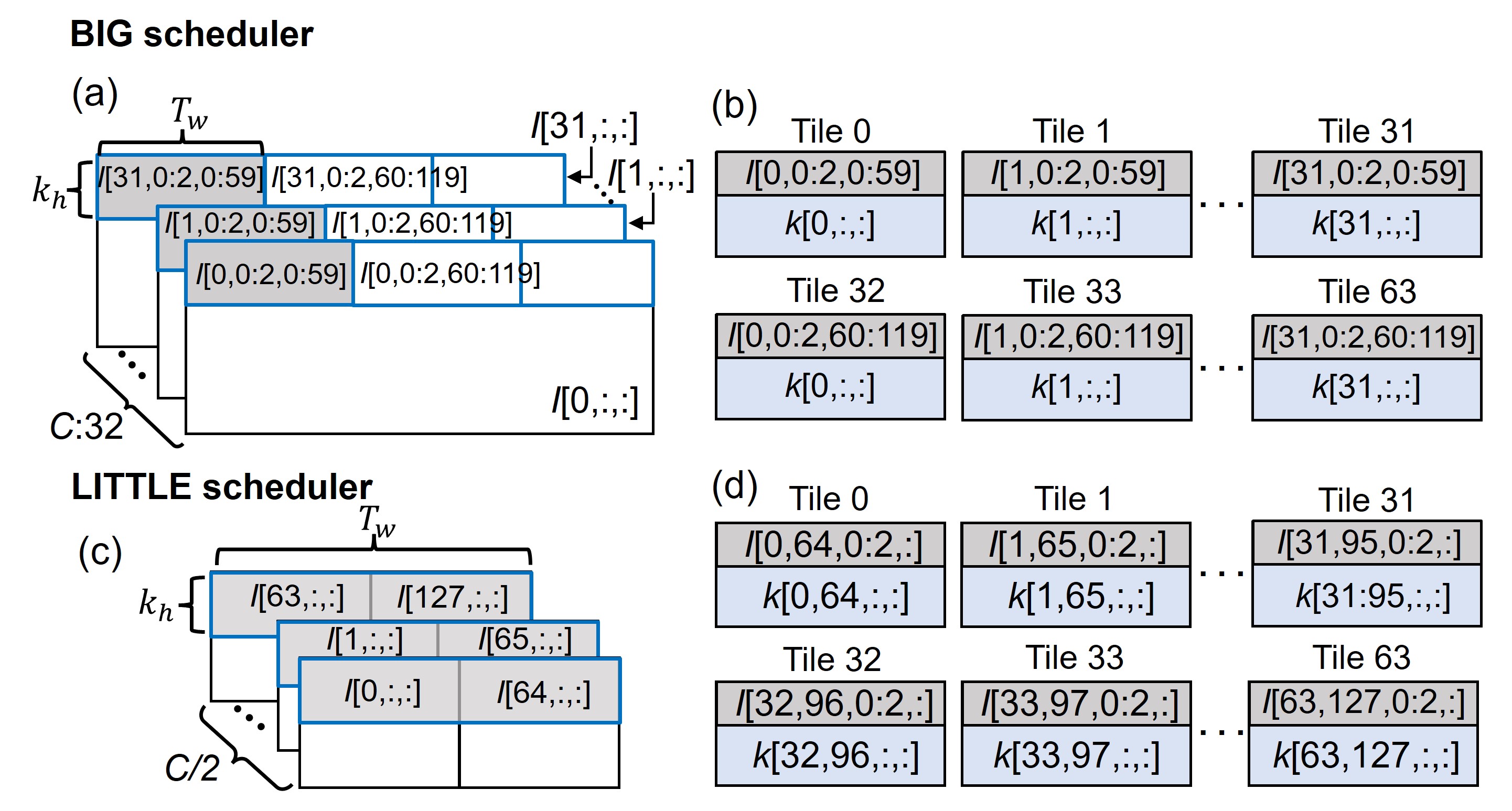}
  \caption{(a) Tiling a 32-channel ifmap using the BIG scheduler for DWConv with a $32\times 3\times 3$ kernel ($k_w=3$ and $T_w=60$) and (b) data mapping across 64 tiles. (c) Tiling a 128-channel ifmap using the LITTLE scheduler for DWConv with a $128\times 3\times 3$ kernel and (d) data mapping across 64 tiles.}
  \label{fig:tiling}
\end{figure}
\textbf{BIG scheduler}: If $W>T_w$ for a $C\times H\times W$ ifmap, the BIG scheduler is selected, which partitions the ifmap into sub-maps, each of size $k_h\times\left[Nk_w+\text{lcm}\left(k_w,s\right)/s-1\right]$, where $N$ is given by Eq.~\eqref{equ:num_dupls}. 
Figs.~\ref{fig:tiling}(a) and (b) illustrate the BIG scheduler for a 32-channel ifmap with a $32\times3\times3$ kernel. The 32 sub-ifmaps across the channels are distributed across Tiles $0-31$. The duplicated 2D kernels for the 32 channels are accordingly distributed over these tiles. 
Using the DWConv principle in Sec.~\ref{sec:datamapping}, each tile performs DWConv in parallel. To increase parallelism and TM utilization, we duplicate the kernel across the 32 idle tiles (Tiles $32-63$) and accordingly distribute next 32 sub-maps across the channel. Generally, the kernel can be duplicated $(\floor*{N_\text{tile}/C}-1)$-times across $N_\text{tile}$ tiles.   

\textbf{LITTLE scheduler}: If $W<T_w$, $W$ determines the duplication number $N$ in Eq.~\eqref{equ:num_dupls}. Particularly, when $W\ll T_w$, the TM utilization for the BIG scheduler is significantly reduced. To maintain high TM utilization, we use the LITTLE scheduler that (1) concatenates partial ifmaps across the channels, (2) partitions the concatenated ifmaps into sub-ifmaps, and (3) distributes them over multiple tiles (Figs.~\ref{fig:tiling}(c) and (d) for a 128-channel ifmap). We define $N_\text{ch}$ as the number of channels hosted by a single tile. In this case, each tile stores a different $N_\text{ch}\times k_h\times k_w$ kernel, and thus performing DWConv for $N_\text{ch}$ channels simultaneously. Similar to the BIG scheduler, the kernels can be duplicated over idle tiles. 

\begin{figure*}[tb]
  \centering
  \includegraphics[width=\linewidth]{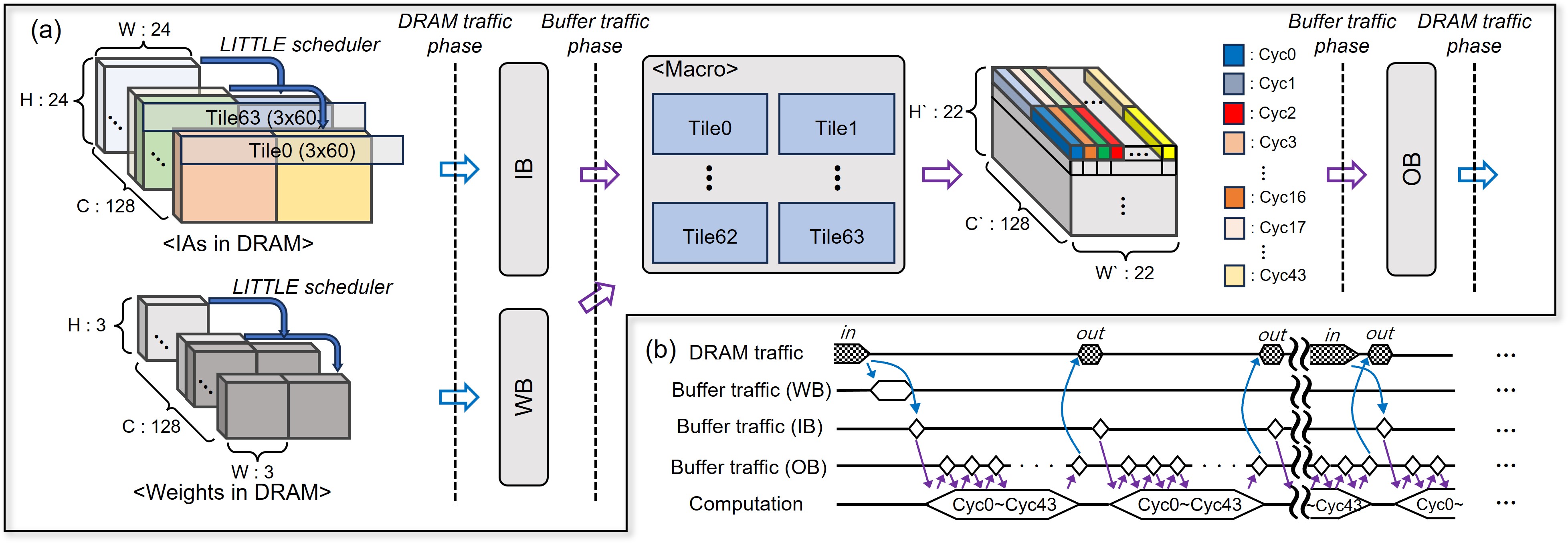}
  \caption{(a) DWConv dataflow using the LITTLE scheduler. A $128\times 24\times 24$ ifmap is subject to DWConv using a $128\times 3\times 3$ kernel ($k_h=k_w=3$) with a stride of 1. Given the $8b\times 180$ TM in each tile, $T_w=60$, and thus $N_\text{ch}=2$. (b) Timing diagram for implementation in the macro.}
  \label{fig:mapping_cycle}
\end{figure*}

\subsection{Dataflow through memory hierarchy for CIM}
Fig.~\ref{fig:mapping_cycle} illustrates the sequence for DWConv using ConvDK with the LITTLE scheduler, including DRAM and buffer traffic. During the DRAM traffic phase, the (partial) ifmap and kernel data are loaded from the DRAM into the 16 kiB IB and 4 kiB WB, respectively. The buffer traffic is primarily attributed to the following three processes: fetching weights into the TM from the WB, fetching partial ifmaps into the TRFs from the IB, and loading the OB with the outputs. 
We designed each $180\times8$b TM to write every kernel element (including duplicates) within a single cycle by simultaneously addressing multiple bitlines at a cycle.
All 64 TMs can be written in parallel. All TRFs are loaded with their corresponding sub-ifmaps from the IB at a single write cycle. 
Once all data are loaded into the macro, DWConv operations are executed in parallel over several compute cycles. This LITTLE scheduler requires $N_\text{ch}H'W'$ compute cycles, where $H'$ and $W'$ are the height and width of the ofmap, respectively. Given \textbf{Algorithm}~\ref{algo:conv_multichannel}, each tile iterates the parallel MAC operations over the $N_\text{ch}(=2)$ channels, so that a cycle for specific $h$ and $w$ outputs $O\left[0:63,h,w\right]$ and the subsequent cycle $O\left[64:127,h,w\right]$. The sub-ofmap is buffered in the OB and then transferred to the DRAM. Note that the DRAM access can be pipelined, causing no additional latency.

\section{Hardware Implementation for ConvDK}
\subsection{CIM operation with 8T-SRAM}
We designed an analog CIM macro performing a bit-wise multiplication using an 8T-SRAM bitcell with two additional pass transistors~\cite{kim2021zpim}. In this structure, multiple 8T-SRAM bitcells share a common read bitline (RBL), which is also connected to a capacitor ($C_\text{RBL}$). For each bitcell, a conductive path between the RBL and GND is established depending on the data stored in the bitcell and input signal. This configuration enables bit-wise multiplication within each bitcell, effectively implementing an AND logic operation. The RBL voltage ($V_\text{RBL}$), initially precharged to $V_\text{ref}$, is discharged only when both the weight bit stored in the bitcell and the bit-serial IA bit, which is applied as the input signal, are `1', and the amount of discharge is proportional to the number of such bitcells. When the two pass transistors operate in the saturation region, $V_\text{RBL}$ can be approximated as
\begin{equation*}\label{equ:discharging}
V_\text{RBL}\approx V_\text{ref}-\frac{1}{C_\text{RBL}}\sum_{i\in IA} (\text{IN}[i]~\text{W}[i]) \cdot I_\text{path} \cdot T_\text{pulse}\text{,}
\end{equation*}
where $\text{IN}[i]$ and $\text{W}[i]$ represent the input signal and stored bit for the \textit{i}-th row, respectively. $I_\text{path}$ is the current through the conductive path formed by the two pass transistors, and $T_\text{pulse}$ denotes the pulse width of the input signal. The set of rows activated simultaneously is denoted as $IA$. The saturation region refers to the operating condition, where the gate-to-source voltage ($V_\text{GS}$) exceeds the threshold voltage ($V_\text{TH}$), and the drain-to-source voltage ($V_\text{DS}$) remains lower than $V_\text{GS}$, allowing the transistor to maintain a nearly constant current and thereby enabling stable and predictable discharging.

\subsection{Multi-access-capable TM for kernel duplication}
In the TM, which holds 180 distributed 8‑bit weights, eight bitcells in each row share a common word line, while all 180 bitcells in each column share the bitline and its complement (bitline bar). Therefore, to store different 8-bit weights, only one word line can be activated (set to HIGH) at a time, requiring 180 clock cycles to write all 180 rows. However, when writing multiple identical data, such as in the case of kernel duplication, multiple word lines can be activated simultaneously, allowing multiple rows to be written in a single clock cycle. For example, when duplicating a $3\times3$ kernel and storing it in the TM, up to 180 weights (including the duplicated ones) can be written in 18 cycles, regardless of the duplication number defined in Eq.~\eqref{equ:num_dupls}. This involves nine cycles for the original weights and one additional cycle per duplicated weight for duplication.

\begin{figure}[tb]
  \centering
  \includegraphics[width=\linewidth]{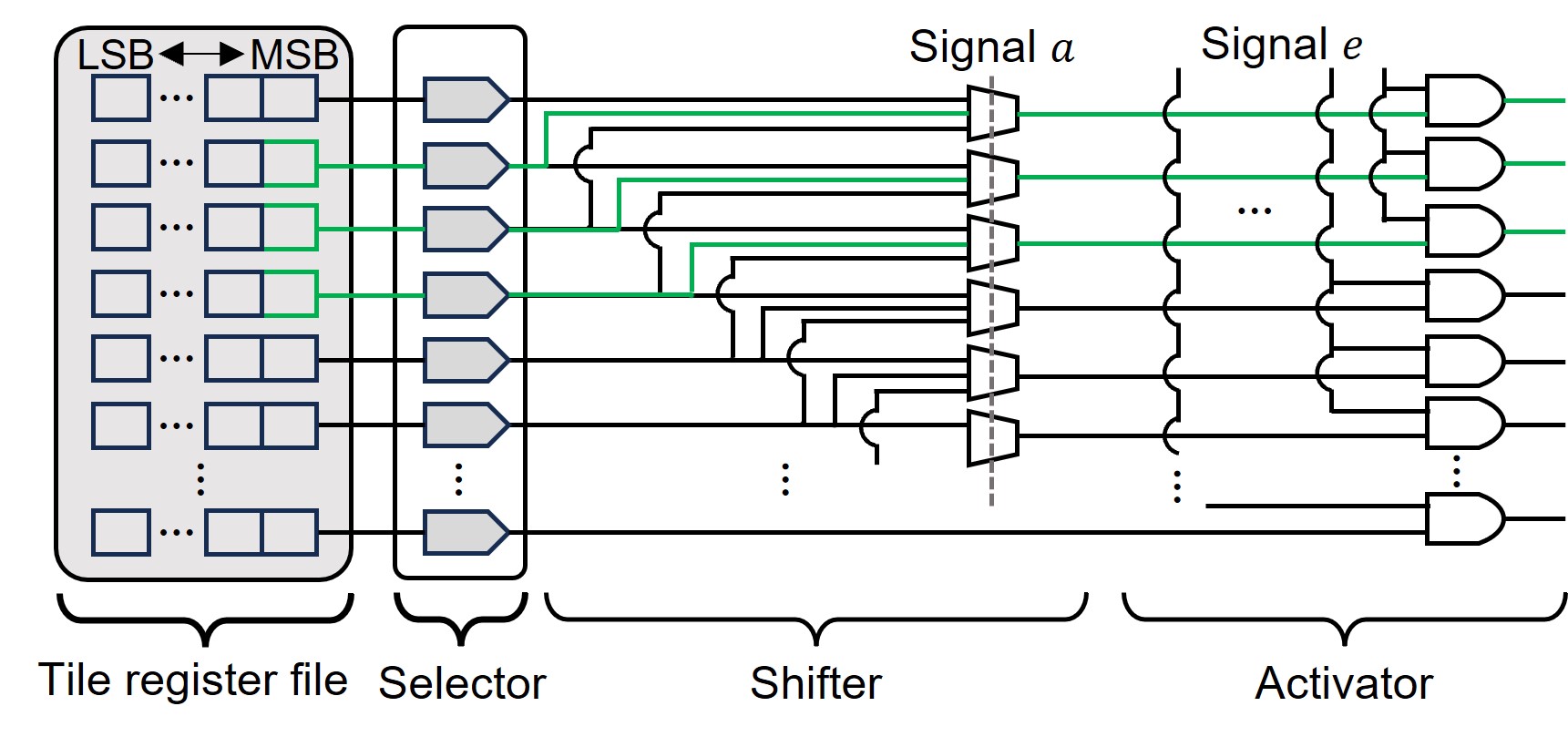}
  \caption{Schematic of the IA shift-and-mask designed for $k_w=3$ and $s=1$, and thus $a\in\left\{0,1,2\right\}$ The multiplication-enable signal is denoted by $e$.}
  \label{fig:architecture}
\end{figure}

\subsection{IA shift-and-mask unit}
We design the IA shift-and-mask (S\&M) unit that enables IA shift by $a$ and blockwise MAC operations controlled by signal $e$ for ConvDK. Fig.~\ref{fig:architecture} illustrates a schematic of IA S\&M for $k_w=3$ and $s=1$, and thus $a\in\left\{0,1,2\right\}$. The selector chooses the digit of IAs in the TRF for MAC operations. The shifter chooses one of three neighboring data using a 3:1 multiplexer controlled by signal $a$. The activator masks all signals from the shifter except a block of signals chosen by multiplication-enable signal $e$ (Sec.~\ref{sec:datamapping}), which are sent to the TM. 

\subsection{Overall latency for three-level memory hierarchy}
This section presents the overall latency, encompassing both computation and data traffic.\\
\textbf{Computation}: The computation latency for parallel MAC operations (up to 16 in parallelism with 8-bit operands) is ten clock cycles, achieved through pipelined execution of the following steps. First, the IA S\&M unit applies IAs to the TM in a bit-serial manner, discharging the $V_\text{RBL}$. The analog accumulation results from the TM are then digitized by the ADC, and the converted digital values are integrated by the S\&A unit. Although this bit-level process is repeated eight times to complete the MAC operation for an 8-bit IA, pipelining across these steps allows the operation to complete within ten clock cycles.\\ 
\textbf{Buffer traffic}: Buffer traffic from the IB to the TRF takes one clock cycle, while the traffic from the WB to the TM takes one clock cycle per 8-bit word. The TRF, directly connected to the buffer via dedicated wires, supports write operations within one clock cycle. In contrast, the SRAM array-based TM performs word-by-word operations, requiring one clock cycle per word for each write. Buffer traffic from the accumulator to the OB also takes one clock cycle.\\
\textbf{Kernel duplication}: Kernel duplication in the multi-access-capable TM incurs an additional latency of one clock cycle.\\
\textbf{DRAM traffic}: Since DRAM traffic operates independently of on-tile computation, data load for upcoming operations (from DRAM to the IB or WB) and store (from the OB to DRAM) can proceed in parallel with the ongoing computation. This decoupling enables pipelined execution of computation and data transfers. Therefore, no additional latency is introduced by the traffic between the DRAM and buffers (IB, WB, and OB) if the data transfer latency is shorter than the computation latency. 
In our work, we considered DDR4-3200, which provides 25.6 GB/s bandwidth at a 1600 MHz clock frequency. This requires 625 ns to fill the 16 KiB IB in our CIM macro~\cite{jedec}. In Fig.~\ref{fig:mapping_cycle}, the computation of all IB data takes $44\times3$ computation cycles, or 5280 ns at a 250 MHz clock frequency (ten clock cycles per one computation cycle), which is the operating frequency used in our simulation. This indicates that the data can be fully loaded from DRAM during the computation period.

\section{Evaluation}\label{sec:eval}
We compared the DWConv performance of our approach (WS ConvDK with the BIG/LITTLE scheduler) with three different dataflows: (1) WS baseline, which is commonly used in CIM macros, (2) IS baseline, which is known to realize the higher utilization for DWConv than WS~\cite{lo2023morphablecim}, and (3) IS ConvDK with the BIG/LITTLE scheduler. We evaluated their performance using five lightweight network models (8-bit IAs and 8-bit weights). The IA S\&M unit, S\&A, and accumulator were synthesized using Synopsys Design Compiler while the rest components were designed using Cadence Virtuoso at a 0.9 V supply voltage and 250 MHz frequency. Table~\ref{tab:hardware_resource} summarizes the area/power overheads for the key components of the CIM macro designed using 28 nm CMOS technology. 

\begin{table}[tb]
\caption{\label{tab:hardware_resource}Summary of memory usage}
\begin{center}
\begin{tabular}{ccccc}
\hline
\multicolumn{5}{c}{\textbf{Memory usage}}\\
\hline
\multicolumn{3}{c}{\textbf{Buffer}} & \multicolumn{2}{c}{\textbf{Tile ($\times 64$)}}\\
\hline
\textbf{IB} & \textbf{OB} & \textbf{WB}& \textbf{TM (8T SRAM)} & \textbf{TRF}\\
\hline
16 kiB & 16 kiB & 4 kiB & 11.25 kiB & 11.25 KiB \\
\hline
\multicolumn{5}{c}{\textbf{Tile ($\times 64$)}}\\
\hline
\multicolumn{2}{c}{\textbf{Module}} &  \multicolumn{2}{c}{\textbf{Area ($\textrm{mm}^2$)}}& \textbf{Power (mW)}\\
\hline
\multicolumn{2}{c}{TM (8T SRAM)} & \multicolumn{2}{c}{0.055} & 0.22 \\
\multicolumn{2}{c}{IA S\&M} & \multicolumn{2}{c}{0.408} & 88.62 \\
\multicolumn{2}{c}{4b-ADC $(\times 8)$} & \multicolumn{2}{c}{0.302} & 36.28 \\
\multicolumn{2}{c}{S\&A} & \multicolumn{2}{c}{0.018} & 3.37 \\ 
\multicolumn{2}{c}{Accumulator} & \multicolumn{2}{c}{0.011} & 2.22 \\
\hline
\end{tabular}
\end{center}
\end{table}

\begin{figure}[tb]
  \centering
  \includegraphics[width=\linewidth]{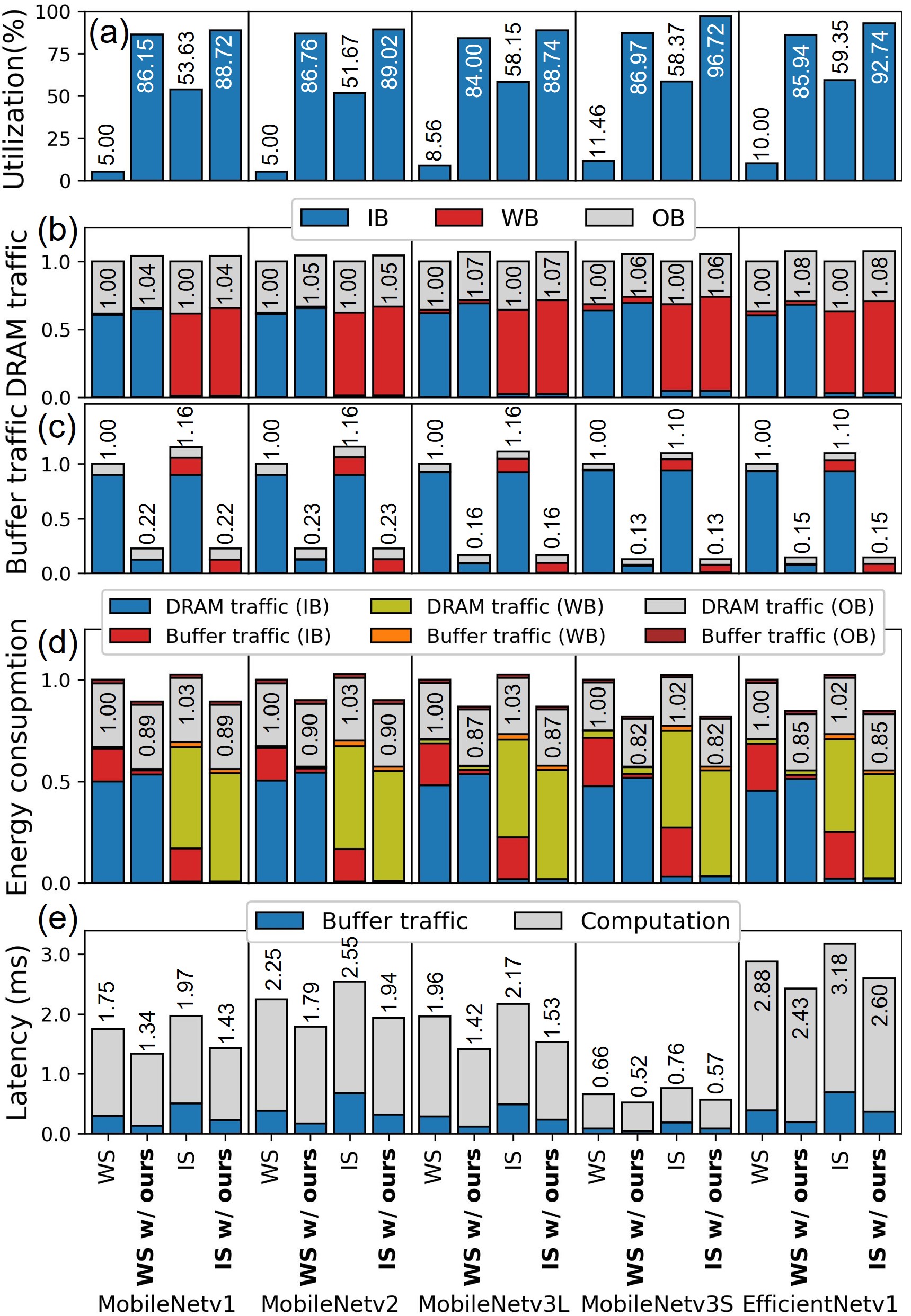}
  \caption{Performance comparison of our dataflow (WS ConvDK with the BIG/LITTLE scheduler with the WS baseline, IS baseline, and IS with our dataflow.)}
  \label{fig:eval}
\end{figure}

\subsection{Utilization}
Fig.~\ref{fig:eval}(a) shows the TM utilization for our dataflows (WS ConvDK with the BIG/LITTLE scheduler) in DWConv-based MobileNetv1/v2/v3L/v3S and EfficientNetv1, which are 86.15\%, 86.76\%, 84.00\%, 86.97\%, and 85.94\%, respectively. While the baseline WS dataflow suffers from significant underutilization of the TM, our method attains high utilization thanks to the weight duplication in ConvDK. Although the IS dataflow notably improves the utilization compared with the baseline, it hardly reaches that for our case because the utilization is constrained by the ifmap size. However, this constraint is largely mitigated by our BIG/LITTLE scheduler, leading to notable increases in the utilization rate. This highlights the versatility of our novel scheduler.

\subsection{DRAM and buffer traffic}\label{sec:access}
Fig.~\ref{fig:eval}(b) presents the DRAM traffic normalized to the baseline. DRAM traffic is primarily determined by the loop-nest order, and buffer level and size~\cite{mei2021zigzag}. Since they are fixed for all four cases, the DRAM traffic is nearly identical across all cases. However, the buffer traffic is primarily determined by IA or weight reuse within the CIM macro. Fig. \ref{fig:eval}(c) highlights our dataflow for WS and IS cases, reducing the buffer traffic by $77.4-87.0\%$ across the five models compared to the WS baseline. This results in significant reductions in energy consumption and latency. 

\subsection{Energy consumption and latency}
We assumed the energy consumption for off-chip DRAM and SRAM-buffer accesses to be 20 pJ/bit~\cite{horowitz2014dramenergy} and 1.139 pJ/bit~\cite{tu2018sramenergy}. The TM and TRF in our design consume 0.017 pJ/bit and 0.028 pJ/bit for writing, respectively.
Fig.~\ref{fig:eval}(d) shows the normalized energy consumed by data traffic within the memory hierarchy including the CIM macro. 
In the WS baseline, the IA movement is dominant, whereas in the IS baseline, the weight movement is dominant, as shown by the red bar in Fig.~\ref{fig:eval}(d). In both cases, our dataflow reduces energy consumption caused by buffer traffic (IB, WB, and OB), achieving reductions of $78.4–87.2\%$ in WS and $81.2–88.3\%$ in IS compared to the baseline. This improvement is mainly attributed to the efficient reuse of buffer data. As a result, the total energy consumption across the entire memory hierarchy, including both DRAM and buffer traffic, is reduced by $10.1–17.9\%$ in WS and $12.8-20.3\%$ in IS compared to the baseline.
This highlights the significant energy cost associated with buffer traffic, which is frequently overlooked.

Finally, Fig.~\ref{fig:eval}(e) shows the latency for all DWConv operations in the five models, taking into account buffer traffic and parallel MAC operations across the 64 tiles. While the compute time occupies the largest portion of the total latency, the buffer traffic latency also occupies a significant portion, ranging from 13.1\% to 16.8\% for the baseline. Note that the DRAM traffic can be pipelined, causing no additional latency. Our dataflow largely reduces the buffer traffic latency.
In the IS cases, the TMs are frequently re-written word-by-word. Although each word and its duplicates are written at a clock cycle, the total write time is larger than that for the TRF which is written at a clock cycle. Therefore, the latency associated with buffer traffic for IS is larger than WS, causing the larger overall latency. 

\begin{figure}[tb]
  \centering
  \includegraphics[width=\linewidth]{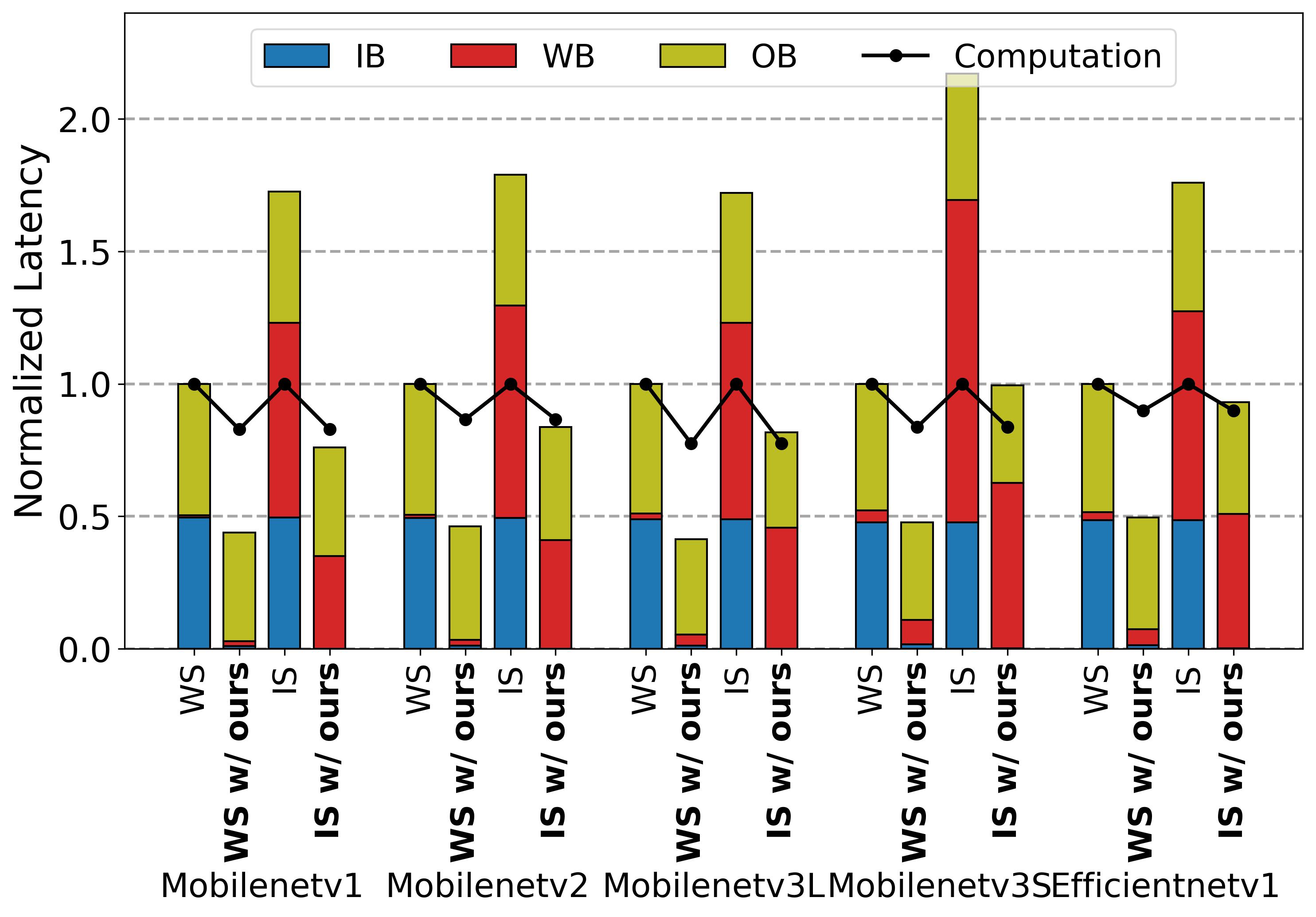}
  \caption{Breakdown of latency caused by buffer traffic. Normalized compute time for each case is also plotted.}
  \label{fig:latency_detail}
\end{figure}

Fig.~\ref{fig:latency_detail} details the breakdown of latency caused by buffer traffic. In the WS cases, our method reduces the buffer traffic by $50.5-58.7\%$. This reduction is primarily attributed to the decrease in traffic between the IB and the TRF. Although, the latency from traffic between the WB and the TM doubles due to kernel duplication, its impact remains negligible because the WB accounts for only $1.0-4.6\%$ of the total buffer traffic latency in the WS baseline. Additionally, the latency associated with the OB decreases by $13.2-26.8\%$, and the computation latency is reduced by $10.1-22.5\%$. Both reductions are attributed to improved TM utilization, facilitated by the BIG scheduler, which enhances MAC parallelism across the 64 tiles. In the IS cases, our ConvDK was also successfully applied, resulting in $47.1-55.9\%$ reduction in buffer traffic. Consequently, the overall latency is reduced by $15.6-27.8\%$ and $18.1-29.3\%$ in the WS and IS cases, respectively, compared to the baseline.

\section{Related work and Comparison with our dataflow}
Morphable CIM~\cite{lo2023morphablecim} is based on an IS systolic dataflow for DWConv with a sub-ifmap placed in the TM. This architecture aims to increase the TM utilization by distributing multiple ifmap channel arrays across the TM columns, which can share weights for DWConv. While this approach effectively enhances data reuse through the systolic operations, the DWConv throughput remains strictly constrained by the number of TM columns working in parallel with parallel ADCs. Additionally, it does not address the latency of writing data in the TM, where data is written word-by-word. As observed in the comparison between the WS and IS cases in Fig.~\ref{fig:eval}(e), the write latency in the IS cases is non-negligible, which may lead to potential performance degradation in the Morphable CIM design that is restricted to the IS dataflow. In contrast, our ConvDK supports both WS and IS dataflows, providing greater flexibility and thus alleviating potential performance degradation.

MobiLattice~\cite{zheng2020mobilatice} addresses the under-utilization issue for DWConv dataflows in resistance-based non-volatile crossbar-based near memory processing. This approach maps a bit-wise vectorized DWConv kernel for a given channel onto a TM column. Each $w$-bit kernel weight is read out by simultaneously addressing $w$ bitlines with different read voltages ($2^{b-1}$ V for $b=1,\cdots,w$) and converting the consequent current sum to the corresponding voltage signal. The voltage signal is subsequently processed by an ADC. The weight is multiplied by IAs using a local processing element. 
While this approach enhances utilization by modifying the weight mapping strategy to improve throughput, it does not emphasize data reuse and therefore offers no benefits in reducing data traffic unlike our proposed ConvDK. From a data movement standpoint, it behaves similarly to the conventional weight-stationary (WS) dataflow. As a result, its performance is comparable to the WS baseline in Fig.~\ref{fig:eval}, assuming identical macro specifications such as the number of ADCs and memory size.

\section{Conclusion}
The proposed ConvDK with the BIG/LITTLE scheduler significantly mitigates the buffer traffic by reusing IAs loaded into the TRFs and enhancing CIM utilization not only within each tile but also across the entire 64 tiles for DWConv operations. The analysis highlights a reduction in the buffer traffic by $77.4-87.0\%$ for five DWConv-based models, and thus reductions in total data traffic energy and latency by $10.1-17.9\%$ and $15.6-27.8\%$, respectively, compared to the baseline (conventional WS dataflow). Additionally, our dataflow successfully applies to IS dataflow, reducing the total data traffic energy and latency by $12.8-20.3\%$ and $18.1-29.3\%$, respectively, compared to the conventional IS dataflow.

\section*{Acknowledgment}
\thanks{This research was supported by Institute of Information \& communications
Technology Planning \& Evaluation (IITP) grants funded by the Korea
government (MSIT) (RS-2023-00229689 and IITP-(2024)-RS-2023-00253914). The EDA tool was supported by the IC Design Education Center (IDEC), Korea.}

\bibliographystyle{IEEEtran}
\bibliography{Bib.bib}

\end{document}